\documentclass[%
 reprint,
 superscriptaddress,
 amsmath,amssymb,
 prapplied, 
]{revtex4-2}
\usepackage[draft]{pgf}
\usepackage{graphicx}
\usepackage{dcolumn}
\usepackage{bm}
\usepackage{amsmath}
\usepackage{gensymb}
\usepackage{tabu}
\usepackage{soul}
\usepackage{mathtools}
\usepackage{epsfig,wrapfig}
\usepackage{epstopdf} 
\usepackage{caption}
\usepackage{subcaption}
\usepackage{fancyhdr}
\usepackage{psfrag}
\usepackage{lipsum}
\usepackage{xcolor}
\usepackage{multirow}
\usepackage{booktabs}
\usepackage{rotfloat}
\usepackage{cancel}
\usepackage[mode=errorstop]{pstool}
\definecolor{cadmiumgreen}{rgb}{0.0, 0.42, 0.24}
\definecolor{darkpink}{rgb}{0.91, 0.33, 0.5}
\definecolor{hotmagenta}{rgb}{1.0, 0.11, 0.81}
\definecolor{fandango}{rgb}{0.71, 0.2, 0.54}
\definecolor{psyblue}{rgb}{0.074, 0.6914, 0.8359}
\usepackage[newcommands]{ragged2e}
\usepackage{graphicx,caption}
\newcommand{\ov}[1]{\overline{#1}}

\renewcommand{\parallel}{\mathrel{/\mkern-5mu/}}
\makeatletter
\newcommand{\notparallel}{%
  \mathrel{\mathpalette\not@parallel\relax}%
}
\newcommand{\not@parallel}[2]{%
  \ooalign{\reflectbox{$\m@th#1\smallsetminus$}\cr\hfil$\m@th#1\parallel$\cr}%
}

\usepackage{hyperref}

\newcommand{\rchi}{\protect\raisebox{2pt}{$\chi$}}

\makeatother

\begin{document}
\preprint{APS/123-QED}

\title{Electrodynamics of Accelerated-Modulation Space-Time Metamaterials}

\author{Amir Bahrami}
\affiliation{
  Department of Electrical Engineering, KU Leuven, Leuven, Belgium
}%

\author{Zo{\'e}-Lise Deck-L{\'e}ger}%
\affiliation{
  Department of Electrical Engineering, Polytechnique Montr{\'e}al, Montr{\'e}al, Quebec, Canada
}%

\author{Christophe Caloz}
\email{christophe.caloz@kuleuven.be}
\affiliation{
  Department of Electrical Engineering, KU Leuven, Leuven, Belgium
}%
\date{\today}

\begin{abstract}
Space-time varying metamaterials based on uniform-velocity modulation have spurred considerable interest over the past decade. We present here the first extensive investigation of \emph{accelerated-modulation} space-time metamaterials. Using the tools of general relativity, we establish their electrodynamic principles and describe their fundamental phenomena, in comparison with the physics of moving-matter media. We show that an electromagnetic beam propagating in an accelerated-modulation metamaterial is bent in its course, which reveals that such a medium curves space-time for light, similarly to gravitation. Finally, we illustrate the vast potential diversity of accelerated-modulation metamaterial by demonstrating related Schwarzschild holes.
\end{abstract}

\maketitle


\clearpage

\section{Introduction} 
Space-Time Modulation~\footnote{The noun ``modulation'' in the term ``space-time modulation metamaterials'' should \emph{not} be replaced by the past participle ``modulated'' because ``modulated metamaterial'' would erroneously suggest that the metamaterial preexists modulation, whereas it is really \emph{formed} by the modulation.} metamaterials are metamaterials that are formed by varying (modulating) some parameter (e.g., the refractive index) of a host medium in both space and time~\cite{cassedy1963dispersion,cassedy1967dispersion,biancalana2007dynamics,hadad2015space,caloz2019spacetime1,caloz2019spacetime2,deck2019uniform,huidobro2021homogenization,galiffi2022archimedes}. The modulated parameter may be of various possible natures, including electronic, optical, acoustic, mechanical, thermal and chemical~\cite{rhodes1981acousto,saleh2019fundamentals,Shaltout_Science_2019}, while the modulation is typically provided by an external drive. These metamaterials are thus \emph{moving-perturbation} media, involving no net transfer of atoms and molecules, and may hence be seen as the modulation counterparts of \emph{moving-matter} media, which are simply referred to as moving media in the literature and whose basic electrodynamics was discovered between the early 19$^\text{th}$ and early 20$^\text{th}$ centuries~\cite{Fresnel1818,Fizeau1851,Rontgen_1888,Minkowski_1908}. Space-time modulation metamaterials share most of the electrodynamics of moving media, including Doppler shifting~\cite{doppler1903ueber}, Bradley aberrations~\cite{bradley1729iv} and light deflection~\cite{einstein1905elektrodynamik}, Fresnel-Fizeau drag~\cite{Fresnel1818,Fizeau1851} and wave-compression amplification~\cite{einstein1905elektrodynamik,yeh1965,granatstein1976}, as shown in a number of studies~\cite{lampe1978interaction,huidobro2019fresnel,caloz2022gstem,xu2022diffusive,tien1958parametric,cullen1960theory}. However, they encompass extra physical regimes, such as instantaneous~\cite{Morgenthaler_1958,halevi2009,halevi2010,hadadsofttemporal,hayranmonticone} and superluminal~\cite{ostrovskiicorrect1967,cassedy1967dispersion,Lurie_Springer_2007,biancalana2007dynamics,Deck_PRB_2018,caloz2019spacetime1,galiffi2021photon,pendry2022crossing} responses, and offer drastic advantages in terms of potential applications, particularly the dispensability of cumbersome moving parts and the easy access to relativistic velocities and accelerations.

The quasi-totality of the research on space-time modulation metamaterials to date has pertained to \emph{uniform-velocity} modulation. Removing the restriction of velocity uniformity by introducing \emph{accelerated-modulation} would naturally imply greater diversity and hence pave the way for novel physics and technology. The related enhancement might perhaps be compared to that gained by extending gravitation-less systems to gravity systems, as done from special relativity~\cite{einstein1905elektrodynamik,d2022introducing} to general relativity~\cite{einstein1915feldgleichungen,misner1974gravity,carroll2019spacetime}, or, according to the principle of equivalence~\cite{einstein1907equivalence}, by extending uniformly moving media to accelerated media. The resulting metamaterials would feature some similarities with conventional gravity analogs~\cite{faccio2013analogue,faccio2012optical} and modulated devices~\cite{rhodes1981acousto,saleh2019fundamentals}, but also transcend them via the incorporation of more sophisticated space-time metrics.

%
\section{Accelerated Space-Time Structures}
\begin{figure}[h!]
 \centering
 \includegraphics[width=0.5\textwidth]{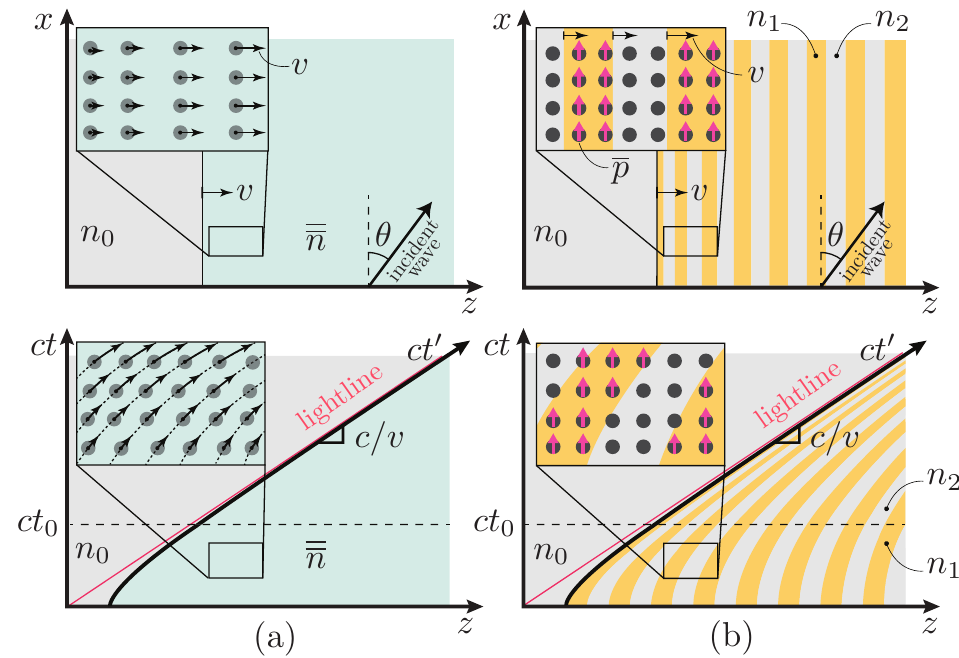}
  \caption{Accelerated- (a)~matter medium and (b)~modulation medium, with constant proper acceleration (Rindler metric) and uniform direction of motion ($z$), in space at time $t=t_0$ (top panels) and in space-time at any position $x$ (bottom panels). The two media are shown as bounded (on their left side) to emphasize their evolution, in a host medium of refractive index $n_0$. The acceleration, corresponding to the constant proper acceleration ($a'$), is inversely proportional to the curvature of the curves (Rindler hyperbolas) in the bottom panels, and is hence time-dependent, $a=a(z,t)$, with $\mathrm{sgn}(a)=\mathrm{sgn}(a')$ [see Appendix~\ref{Sec:GR_Tools}].}
  \label{fig:physics} 
\end{figure}

Figure~\ref{fig:physics} shows the space (top panels) and space-time (bottom panels) structures of the accelerated-matter and accelerated-modulation media. It specifically corresponds to the regime of \emph{constant proper acceleration} and \emph{uniform direction of motion}, which will pertain to most of the paper. This regime is chosen because it is the simplest possible acceleration regime while  sufficing to reveal the most fundamental physics of accelerated-modulation space-time  metamaterials. The last part of the paper, on analog holes, will involve a more complex acceleration regime, featuring \emph{varying acceleration} and \emph{nonuniform, spherical direction of motion}, but featuring a structure that is locally similar to that in Fig.~\ref{fig:physics}(b) and basic principles that draw from the related physics.

Figure~\ref{fig:physics}(a) represents the structure of an accelerated-matter medium. Such a medium is typically an object that has been propelled by a mechanical force, and the moving entities are the atoms and molecules [shown in the insets of Fig.~\ref{fig:physics}(a)] that form the object. The motion of these entities induces magnetoelectric coupling~\cite{Rontgen_1888}, which generally makes the medium bianisotropic, with well-known tensorial permittivity, permeability and coupling parameters, $\ov{\ov{\epsilon}}$, $\ov{\ov{\mu}}$ and $\ov{\ov{\rchi}}$~\cite{kong1990electromagnetic}.

Figure~\ref{fig:physics}(b) depicts the structure of an accelerated-modulation medium. Such a medium may be formed, for instance, by illuminating a dielectric/piezoelectric slab with a circular phase-front optical/acoustic pump, or by switching a varactor-loaded artificial transmission line structure according to a nonuniform voltage sequence. Here, the atoms and molecules that form the structure [shown in the insets Fig.~\ref{fig:physics}(b)] do \emph{not} move; what moves is only the modulation, which is a traveling-wave, typically sinusoidal, perturbation of the refractive index. \textcolor{black}{We shall assume throughout the paper that the structure is operating in the \emph{metamaterial regime}, where the modulation is subwavelength ($\lambda_\mathrm{modulation}\ll \lambda_\mathrm{wave}$) and subperiod ($T_\mathrm{modulation}\ll T_\mathrm{wave}$)~\cite{huidobro2021homogenization,huidobro2019fresnel,caloz2022gstem}, so that the striped structure in Fig.~\ref{fig:physics}(b) can be reduced, by proper averaging, to a \emph{homogeneous medium}~\footnote{\color{black}{The homogeneous metamaterial regime is the long-wavelength and long-period (viewpoint of the wave) regime that is located well below spatial and temporal frequencies of the Bragg limit~\cite{deck2019uniform} beyond which the stopband structure starts, where only the lowest s-polarized mode (considered in the paper) and the lowest  p-polarized mode (not considered in the paper) exist in the medium}\color{black}}, as will be done in Sec.~\ref{sec:acc_mod_med}}. In this regime, the sinusoidal modulation profile may be approximated by a bilayer periodic profile, with layer indices $n_1$ and $n_2$, and the corresponding configuration may be seen as an effective ``atom - vacuum'' sequence that is akin to the atom-vacuum structure of the real medium in Fig.~\ref{fig:physics}(a).

\section{Methodology and Assumptions}
We shall now derive and compare the dispersion relations and Poynting-vector directions for the two media in Fig.~\ref{fig:physics}. This will be done using some mathematical tools of general relativity, specifically the Maxwell-Cartan equations~\cite{cartan1924varietes}, the Rindler transformations and metric~\cite{rindler1960hyperbolic}, and the general tensorial coordinate transformation formulas~\cite{tu2017differential} (see Appendix~\ref{Sec:GR_Tools}), and making extensive use of frame hopping between the laboratory frame, $K$ (unprimed variables), and the moving frame, $K'$ (primed variables). We shall assume that the problem is invariant in the $y$ direction (Fig.~\ref{fig:physics}), and hence spatially two-dimensional, and restrict our attention to the case of s polarization\textcolor{black}{, where the electric field is directed along the $y$ direction}, the case of p-polarization being formally analogous.
 
\section{Accelerated-Matter Medium}
Let us start our electrodynamic study with the case of the accelerated-matter medium [Fig.~\ref{fig:physics}(a)]. As is usually done in such problems, to avoid the complexity of bianisotropy, we first compute the dispersion relations and the fields in the $K'$ frame, whose noninertiality requires a general-relativistic treatment, and then transform these quantities back to the $K$ frame. Writing the $K'$-frame dispersion relation in its covariant form, i.e., $g_{\mu'\nu'}k^{\mu'}k^{\nu'}=0$, where $k^{\mu'}=(\omega'/c,\mathbf{k'})$, and inverse-Rindler transforming the resulting relation yields the $K$-frame dispersion relation (see Appendix~\ref{Sec:Mat_Disp})
\begin{equation} \label{eq:Dispersion_Matter_mainT}
	 \frac{\left(k_z-\rchi \omega/c\right)^2}{\alpha^2 n'^2}+\frac{k_x^2}{\alpha n'^2}=\left(\frac{\omega}{c}\right)^2.
\end{equation}
In this relation, $n'$ is the ($K'$-frame) refractive index of the medium, which is assumed to be isotropic, dispersionless and linear so that $n'$ is a constant, $\alpha=(1-\beta^2)/(1-n'^2\beta^2)$ and $\rchi=\beta(1-n'^2)/(1-n'^2 \beta^2)$, where $\beta=v/c=\tanh(a't'/c+\xi_0)$ is the instantaneous normalized velocity of the medium [Fig.~\ref{fig:physics}(a)], with $a'$ being the (constant) proper acceleration and $\xi_0=\sinh^{-1}(\beta_0\gamma_0)$, where $\beta_0=v(0)/c$ and $\smash{\gamma_0=1/\sqrt{1-\beta_0^2}}$ are the normalized initial velocity and the corresponding Lorentz factor. 

On the other hand, the direction of the Poynting vector, $\mathbf{S}$, \textcolor{black}{may be obtained by} writing the wave equation in the $K'$-frame, solving it using a vacuum-field ansatz, and inverse-Rindler transforming the resulting field expressions~\cite{tanaka1978relativistic}, which yields (see Appendix~\ref{Sec:Mat_Fields})
\begin{equation}\label{eq:Angle_Poynting_Mat_iso}       
 \theta_\mathbf{S}=\tan^{-1}\left(\gamma\frac{\sin\theta_n'+n'\beta}{\cos\theta_n'}\right),
\end{equation}
where $\smash{\gamma=1/\sqrt{1-\beta^2}=\cosh(a't'/c+\xi_0)}$ is the instantaneous Lorentz factor and $\sin\theta_n'=N/D$ with $N=1-\sin\theta\tanh(a't'/c n'+\xi_0)$ and $D=\sin \theta-\tanh (a't'/cn'+\xi_0)$, where $\theta$ is the incident (initial) angle\textcolor{black}{, shown in the top panels of Fig.~\ref{fig:physics}}. Note that Eq.~\eqref{eq:Angle_Poynting_Mat_iso}, although partially expressed in terms of $K'$-frame variables, for the sake of compactness, really represents a $K$-frame quantity.

\section{Accelerated-Modulation Medium}\label{sec:acc_mod_med}

Let us move on now to the case of the accelerated-modulation medium, which is the main object of this paper. The corresponding analysis is considerably more involved than that of the accelerated-matter medium, due to both the more complex, bilayer structure of the medium [Fig.~\ref{fig:physics}(b)] and the inexistence of a motion-less frame, as we shall see next. Here, in the $K$ frame, the modulation is accelerating and matter is stationary, while in the $K'$ frame, the modulation is stationary and matter is accelerating, in the opposite direction, with velocity $v'=-v$. So, motion occurs in both frames, which is unusual in conventional relativity problems. We still elect to attack the problem in the $K'$ frame, on the ground that moving matter with stationary boundaries [here, the interfaces between the layers in Fig.~\ref{fig:physics}(b)] is a known problem~\cite{deck2021electromagnetic}, where the addition of noninertiality is tractable using the tools of general relativity.

In the $K'$ frame, the two media forming the stratified structure [Fig.~\ref{fig:physics}(b)], assumed to be isotropic in $K$, with scalar refractive indices, relative permittivities and relative permeabilities $n_{1,2}$, $\epsilon_{1,2}$ and $\mu_{1,2}$ (\smash{$n_{1,2}=\sqrt{\epsilon_{1,2}\mu_{1,2}}$}), are bianisotropic, due to the motion of their constituent matter, and noninertial, due to acceleration. Upon this consideration, the sought after dispersion relation may be obtained by first constructing the corresponding $K'$-frame constitutive relations, which include the tensorial constitutive parameters $\smash{\ov{\ov{\epsilon}}_{1,2}'}$, $\smash{\ov{\ov{\mu}}_{1,2}'}$ and $\smash{\ov{\ov{\chi}}_{1,2}'}$, then space-time averaging the tangential and normal components of these parameters (metamaterial regime), next inverse-Rindler transforming the so-obtained averages, and finally substituting the resulting expressions into the general ($K$-frame) bianistropic dispersion relation~\cite{kong1990electromagnetic} (see Appendix~\ref{Sec:Mod_Disp}). This yields 
\begin{equation}\label{eq:Dispersion_Modulation_main}
	 \frac{\left( k_z-\ov{\rchi} \omega/ c\right)^2}{\ov\epsilon_\|\ov\mu_\|}+\frac{k_x^2}{\ov\epsilon_\|\ov\mu_\perp}=\left(\frac{\omega}{c}\right)^2,
\end{equation}
which involves the averaged tangential and normal constitutive parameters
\begin{subequations}\label{eq:Avg_Params}
\begin{equation}
    \ov\epsilon_\|=\frac{\Sigma_\epsilon-\beta^2\epsilon_1\epsilon_2\Sigma_\mu}{1-\beta^2\Sigma_\epsilon\Sigma_\mu},\quad\ov\epsilon_\perp=\frac{2\epsilon_1\epsilon_2}{\Sigma_\epsilon},
\end{equation}
\begin{equation}
    \ov\mu_\|=\frac{\Sigma_\mu-\beta^2\mu_1\mu_2\Sigma_\epsilon}{1-\beta^2\Sigma_\epsilon\Sigma_\mu},\quad \ov\mu_\perp=\frac{2\mu_1\mu_2}{\Sigma_\mu}
\end{equation}
and
\begin{equation}\label{eq:chibmod}
    \ov{\rchi}=\beta\frac{\Delta_\epsilon\Delta_\mu}{1-\beta^2\Sigma_\epsilon\Sigma_\mu},
\end{equation}
where $\Sigma_\epsilon=(\epsilon_1+\epsilon_2)/2$, $\Delta_{\epsilon}=(\epsilon_1-\epsilon_2)/2$, and similarly for $\Sigma_\mu$ and $\Delta_\mu$.
\end{subequations}

On the other hand, the Poynting vector direction may be obtained by taking the spatial Fourier transforms of the $K$-frame Maxwell equations and dispersion relations, inserting the resulting former expression into the resulting latter expression, eliminating the electric and magnetic flux density fields, and taking the appropriate ratio of the remaining field components (see Appendix~\ref{Sec:Mod_Defl}), which yields
\begin{equation} \label{eq:Poynting_Angle}
   \theta_{\mathbf{S}}=\tan^{-1}\left({\frac{\ov\mu_\perp}{\ov\mu_\|}\frac{\sin\theta-\ov{\rchi}}{\cos\theta}} \right),
\end{equation}
whose parameters are given by Eqs.~\eqref{eq:Avg_Params}.

\section{Formal Modulation-Matter Equivalence}
Comparing Eq.~\eqref{eq:Dispersion_Modulation_main} with Eq.~\eqref{eq:Dispersion_Matter_mainT} reveals a most interesting fact: the accelerated-modulation medium is formally equivalent to the accelerated-matter medium (in $K$). Both media are indeed bianisotropic, with identical tensorial structure. Thus, an accelerated-modulation space-time  metamaterial can potentially alter light in the same manner as accelerated matter and hence, by the equivalence principle, as gravitation.

\section{Dispersion and Propagation}
\begin{figure}[!h] 
\centering
\includegraphics[width=0.5\textwidth]{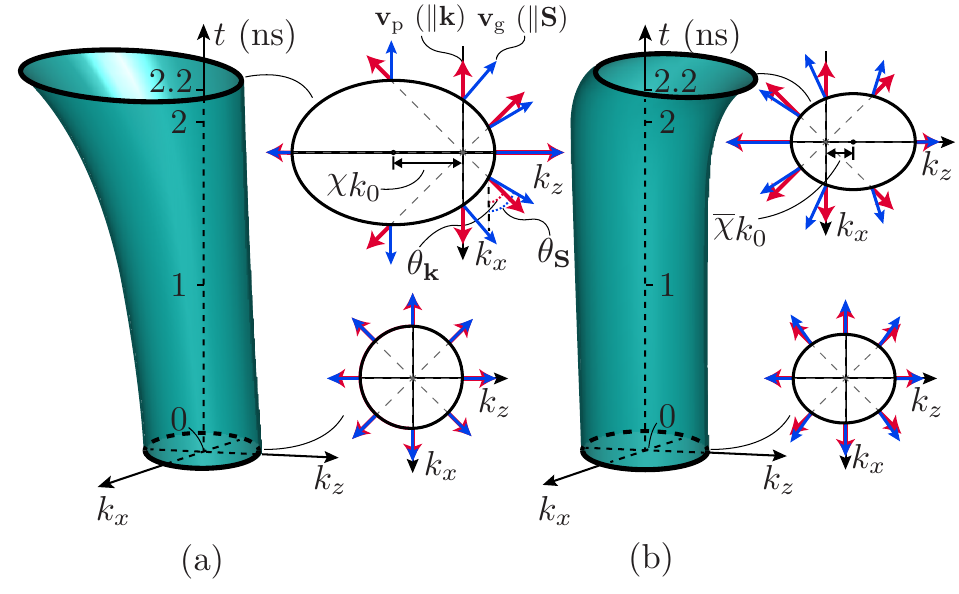}
\caption{Temporal evolution of isofrequency curves for the accelerated-(a) matter medium [Eq.~\eqref{eq:Dispersion_Matter_mainT}] and~(b) modulation medium [Eq.~\eqref{eq:Dispersion_Modulation_main} with~\eqref{eq:Avg_Params}] in Fig.~\ref{fig:physics}, with proper acceleration $a'/c^2=0.45$ and initial normalized velocity $\beta_0=0$. The time $t=0$ corresponds to the onset of the modulation. The modulation medium has the $K$-frame refractive indices $n_1=\sqrt{\epsilon_1\mu_1}$ and $n_2=\sqrt{\epsilon_2\mu_2}$, with $\epsilon_1=\mu_1=1.5$ and $\epsilon_2=\mu_2=3$, while the matter medium has the $K'$-frame refractive index $n'=\sqrt{\epsilon'\mu'}$ with $\smash{\epsilon'=\Sigma_\epsilon}$ and $\smash{\mu'=\Sigma_\mu}$, corresponding to $n'=2.25$. The insets show the isofrequencies $t=0$ and $t=2.2$~ns and corresponding representative phase velocity vector, $\mathbf{v}_\mathrm{p}$, and group velocity vector, $\mathbf{v}_\mathrm{g}$ [with directions given by Eqs.~\eqref{eq:Angle_Poynting_Mat_iso} and~\eqref{eq:Poynting_Angle}].}
 \label{fig:Dispersion}
\end{figure}

\begin{figure}[h!] 
   \centering
  \includegraphics[width=0.5\textwidth]{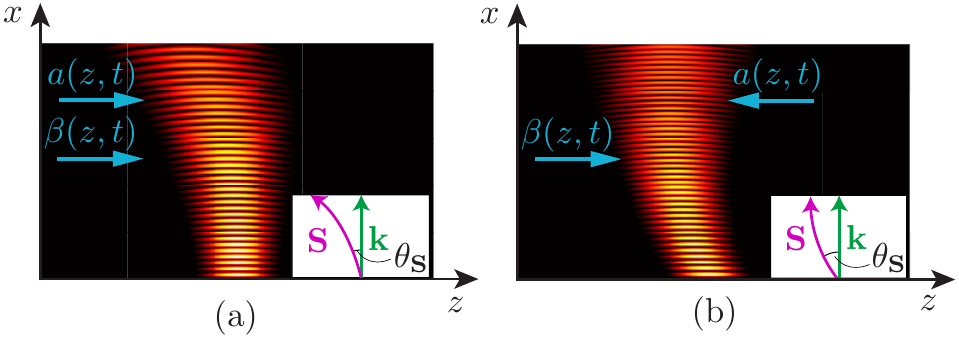}
  \caption{Deflection and bending of a Gaussian beam in the accelerated-modulation medium in Fig.~\ref{fig:physics}(b), specifically $\left|\mathrm{Re}\left\{E_y\right\}\right|$ [Eq.~\eqref{eq:E_final}], with the medium parameters in Fig.~\ref{fig:Dispersion} ($\epsilon_1=\mu_1=1.5$ and $\epsilon_2=\mu_2=3$), $W_0=1.5\lambda_0=0.45$~mm and $|a'|/c^2=0.45$, for the cases of (a)~positive acceleration, with $\beta_0=0$, and (b)~negative acceleration (or deceleration), with $\beta_0=0.3$. \textcolor{black}{The arrows associated with $a$ and $\beta$ correspond to the direction of the acceleration and velocity, respectively}.}
  \label{fig:Analt_Deflection}
\end{figure}

Figure~\ref{fig:Dispersion} shows typical temporal evolutions of the isofrequency contours for the accelerated media of interest. Figure~\ref{fig:Dispersion}(a) corresponds to the accelerated-matter case, computed by Eq.~\eqref{eq:Dispersion_Matter_mainT}, while Fig.~\ref{fig:Dispersion}(b) corresponds to the accelerated-modulation case, computed by Eq.~\eqref{eq:Dispersion_Modulation_main}, respectively. As time passes, the isofrequency contours progressively shift, in both cases, parallel to the direction of motion, $z$. In the case of accelerated matter [Fig.~\ref{fig:Dispersion}(a)], the shift occurs in the $-z$ direction, which corresponds to deflection of energy (direction $\mathbf{v}_\mathrm{g}$ in the figure) in the $+z$ direction. This deflection is a manifestation of the Fresnel-Fizeau drag~\cite{Fresnel1818,Fizeau1851}, which adds the momentum $\ov{\rchi} k_0\propto\beta$, increasing with time due acceleration, to the $k_z$ component of the wave [Eq.~\eqref{eq:Dispersion_Matter_mainT}]. The physics is drastically different in the case of accelerated modulation [Fig.~\ref{fig:Dispersion}(b)]. Now, the isofrequency contour shift is in the $+z$ direction, corresponding to deflection of energy in the $-z$ direction, i.e., \emph{contra-directionally} to the motion~\cite{deck2019uniform,huidobro2019fresnel,caloz2022gstem}. Such contra-directional deflection is \emph{not} due to the Fresnel-Fizeau effect, since no motion of matter occurs in the laboratory frame; it results from the \emph{space-time weighted averaging}~\cite{deck2019uniform} of the constitutive parameters in Eq.~\eqref{eq:Avg_Params}, whereby the wave traveling along the direction of the modulation, spending more time in the denser layers ($n_1$, assuming $n_1>n_2$), propagates slower in that direction than in the opposite direction~\cite{caloz2022gstem}
\textcolor{black}{\footnote{\textcolor{black}{This space-time weighted averaging effect was discovered in~\cite{deck2019uniform} and extensively described in~\cite{caloz2022gstem}. In these papers, the effect was derived for the case of a \emph{uniformly moving}  modulation, but the related argument still applies to the case of acceleration because an accelerated medium may be considered as having locally a uniform velocity at any given point of space and time [e.g., final time in the inset of Fig.~\ref{fig:Dispersion}(b)], so that the global deflection of the wave energy (global direction of the beam in Fig.~\ref{fig:Analt_Deflection}) is entirely determined by a single point of space-time. The deflection of a beam in a uniform-velocity modulation medium was first of explicitly demonstrated in~\cite{huidobro2019fresnel}, but with the problematic invocation of the ``Fresnel drag'' as an explanation for the effect. The Fresnel-Fizeau~\cite{Fresnel1818,Fizeau1851} drag implies moving matter (fluid in Fizeau's original experiment~\cite{Fizeau1851}), whereas no matter (atoms and molecules) is moving in the moving-modulation medium. One could argue that interfaces move, and that the ``Fresnel drag'' would be one related by these moving interfaces instead of moving matter, but the problem is that this would predict the wrong direction of deflection!}}}, which imparts the negative momentum contribution $-\left|\ov{\rchi}\right|k_0\propto-\beta$ to the $k_z$ component of the wave [Eq.~\eqref{eq:Dispersion_Modulation_main}]~\footnote{Note that the isofrequency curve at $t=0$ (before the onset of the modulation) is \emph{not} a circle [Fig.~\ref{fig:Dispersion}(b)]. Mathematically, this may be seen by inserting Eq~\eqref{eq:Avg_Params} into~\eqref{eq:Dispersion_Modulation_main} with $\beta=0$, which implies $\ov{\rchi}=0$; this operation centers the curve to the origin (absence of motion) but does not suppress its eccentricity. The remaining elliptical shape simply corresponds to the anisotropy of the stratified structure of the modulation medium [see Fig.~\ref{fig:physics}(b) at $t=0$].}.

In order to further examine this contra-directional deflection effect, let us consider the propagation of a Gaussian beam, launched perpendicularly to the direction of motion [$x$ direction in Fig.~\ref{fig:physics}(b)] for maximal deflection. Such a field may be obtained by inserting the paraxial approximation of the dispersion relation in Eq.~\eqref{eq:Dispersion_Modulation_main} into the integral expression for the field corresponding to the propagating Gaussian beam, and calculating the resulting integral, which yields (see Appendix~\ref{Sec:Mod_Fields})
\begin{equation}\label{eq:E_final}
\begin{split}
    E_y(z,x)=E_0\frac{\text{e}^{\mathrm{i}n_\| k_0(1-\delta)x}}{1+\mathrm{i}\frac{n_\|2x}{n_\perp^2k_0W_0^2}}
    \exp\left(-\frac{\left(z+z_{\delta}\right)^2}{\left(1+\mathrm{i}\frac{n_\|2x}{n_\perp^2k_0W_0^2}\right)W_0^2}\right),
\end{split}
\end{equation}
where $\smash{\ov n_\|=\sqrt{\ov\epsilon_\|\ov\mu_\perp}}$, $\smash{\ov n_\perp=\sqrt{\ov\epsilon_\|\ov\mu_\|}}$, $\smash{\delta=\ov{\rchi}^2k_0/2\ov n_\perp^2}$ and $\smash{z_\delta=x\ov{\rchi}\ov n_\|/\ov n_\perp^2}$ [Eqs.~\eqref{eq:Avg_Params}], with $k_0=\omega/c=2\pi/\lambda_0$, and where $W_0$ is the waist of the beam at $x=0$~\footnote{Note that all the formulas given in the paper are complicated functions of the odd parameter $\beta$ such that the related quantities and relations are nonreciprocal.}. 

Figure~\ref{fig:Analt_Deflection} plots the Gaussian-beam field, computed by Eq.~\eqref{eq:E_final}, with Figs.~\ref{fig:Analt_Deflection}(a) and~\ref{fig:Analt_Deflection}(b) corresponding to positive modulation acceleration and negative modulation acceleration (or deceleration), respectively. A first observation is that the beam, in both cases, is deflected away from the launching axis ($x$), i.e., in the opposite direction of the velocity ($\beta$), as expected from the previous results [Fig.~\ref{fig:Dispersion}(b)]. However, the figure also reveals that the beam is \emph{bent} by the medium, with bending occurring \emph{towards the direction of the acceleration}~($a$). This result indicates that \emph{accelerated modulation curves space-time} for light, similarly to gravitation~\cite{einstein1914formale,einstein1915feldgleichungen,misner1974gravity}. In fact, such curving could have been inferred from the (known) existence of deflection (without curving) in a uniform-velocity (or acceleration-less) modulation 
 medium~\cite{deck2019uniform,huidobro2019fresnel,caloz2022gstem} [Eq.~\eqref{eq:Dispersion_Modulation_main} with $\ov{\rchi}=\mathrm{const.}$ from $\beta(a'=0)=\beta_0=\mathrm{const.}$, corresponding to a straight, vertical elliptic cylinder in Fig.~\ref{fig:Dispersion}(b)]. Indeed, since acceleration is locally equivalent to uniform velocity and since uniform-velocity sections with different velocities at different positions of space-time point to different directions, combining the related infinitesimal uniform sections automatically produces the observed curving. However, the exact dispersion and field quantities given above could naturally not have been obtained without a rigorous general relativistic treatment.

\section{Light Bending for Various Constituent Media}
\begin{figure}[!h] 
    \centering
  \includegraphics[width=0.5\textwidth]{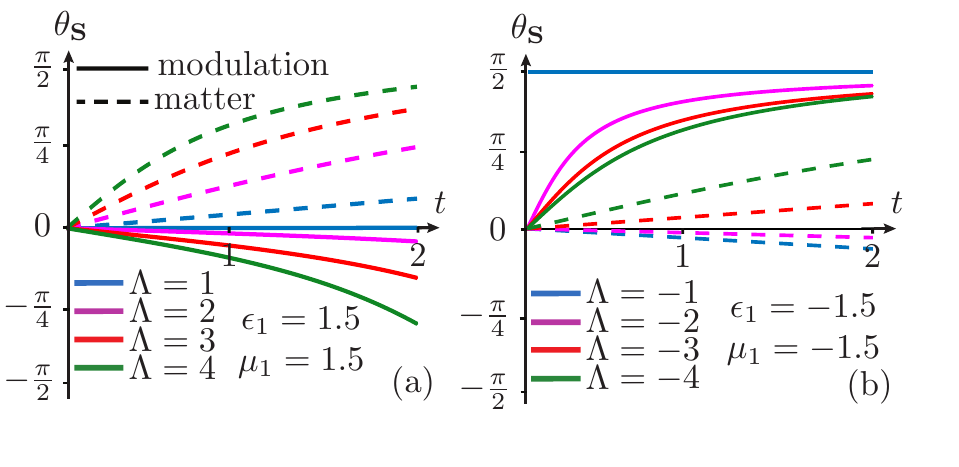}
      \caption{Direction of the Poynting vector (angle shown in Figs.~\ref{fig:Dispersion} and~\ref{fig:Analt_Deflection}) versus time for different values of the parameter \mbox{$\Lambda=\epsilon_2/\epsilon_1=\mu_2/\mu_1$} (with $\smash{\sqrt{\mu_1/\epsilon_1}=\sqrt{\mu_2/\epsilon_2}}$ for local matching) in the accelerated-modulation medium [Eq.~\eqref{eq:Poynting_Angle}] and in the accelerated-matter medium [Eq.~\eqref{eq:Angle_Poynting_Mat_iso}] for \mbox{$a'/c^2=0.83$} (and hence $a>0$) and $\beta_0=0$ (and hence $v>0$). The index $n'$ for matter has been chosen as in Fig.~\ref{fig:Dispersion} for proper Fresnel-Fizeau density and quantitative comparison. (a)~Double-positive ($n'^2>0$) and (b)~double-negative ($n'^2>0$) medium parameters.}
  \label{fig:Deflection_Comparison}
\end{figure}

So far, we have implicitly assumed that the two media forming the unit cell of the space-time modulation metamaterial in Fig.~\ref{fig:physics}(b) have specific double-positive constitutive parameters  (\mbox{$\epsilon_1,\mu_1>1$} and $\epsilon_2,\mu_2>1$). However, modulation parameters might take various and even negative values. Therefore, we shall now perform a parametric analysis of the light bending effect for different constitutive parameters, and compare the results with those of moving matter with equivalent average parameters. This analysis is presented in Fig.~\ref{fig:Deflection_Comparison}, which plots the temporal evolution of the Poynting vector direction for a wave launched in the $x$ direction, with Figs.~\ref{fig:Deflection_Comparison}(a) and (b) respectively corresponding to different double-positive and double-negative medium parameters.

In the case of matter, all the results in Fig.~\ref{fig:Deflection_Comparison} (dashed curves) may be simply explained in terms of the Fresnel-Fizeau drag effect, according to which the velocity of light in the moving-matter medium is given by \mbox{$v_{\text{light}}=c/n'+v(1-1/n'^2)$}~\cite{Fizeau1851}, where $n'$ is the refractive index of the matter medium at rest. Since it occurs in the $z$ direction, the drag effect implies $\theta_\mathbf{S}\propto v(1-1/n'^2)$. It may be easily checked that results in the figure precisely follow this prediction for the two different types of medium constituents, with the factor $v$ ($\partial v/\partial t=a>0$) accounting for the temporal evolution of any given curve and factor $1-1/n'^2$ accounting the signs and difference between the different curves at any given time. 

Let us now examine the results for the modulation case (solid curves in Fig.~\ref{fig:Deflection_Comparison}). The figure shows that the metamaterial can reach a great range of energy deflection and bending amounts upon proper parametric tuning, similarly to its matter medium counterpart, as expected from the formal equivalence between the two types of media that was previously pointed out. These results are harder to explain than those for matter, due to the greater complexity of the metamaterial structure and related space-time averaged quantities [Eqs.~\eqref{eq:Avg_Params}]. However, the negative direction of deflection and bending for the double-positive constituent-media structure [Figs.~\ref{fig:Deflection_Comparison}(a)] clearly corresponds to the motion-contradirectional effect that was explained in connection with Fig.~\ref{fig:Dispersion}(b), while the strength of the effect is explained by the amount of contrast between the layer parameters, $\Delta_{\epsilon,\mu}$, since it is this contrast that forms the medium. Moreover, the results for the double-negative constituent-media structure [Fig.~\ref{fig:Deflection_Comparison}(b)] may be understood as follows. For $\Lambda=-1$, the structure alternates equal-magnitude positive and negative index layers so as to form a Pendry lens~\cite{pendry2000lense} periodic configuration~\cite{kong2002stratified}, here with moving interfaces, where the positive and negative $x$ momentum components across the unit cell cancel out so as to produce purely $z$-directed propagation. For $\Lambda\neq-1$, the refractive indices of the two layers of the unit cell have different magnitudes, which breaks the $\Lambda=-1$ $x$-momentum anti-symmetry; as a result, the rays in the two layers undergo different deflections, which produces a net wave deflection, proportional to the index contrast.

\section{Analog-Gravity Media}
\begin{figure}[h] 
   \centering
  \includegraphics[width=0.5\textwidth]{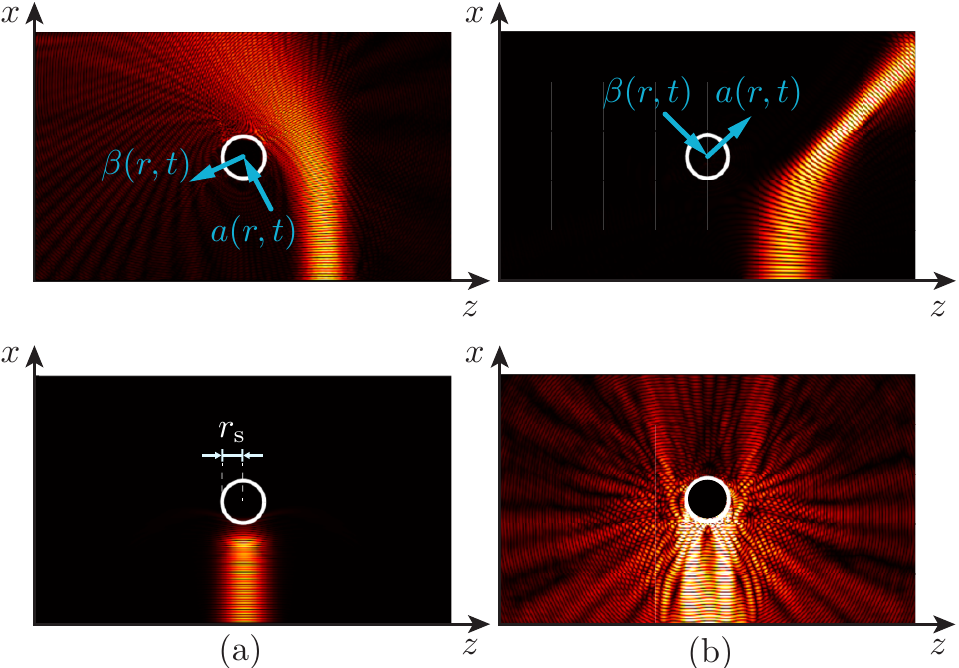}
  \caption{Scattering of a Gaussian beam of light by Schwarzschild (a)~black and (b)~white holes based on a cylindrical version of the accelerated bilayer-unit-cell space-time modulation medium in Fig.~\ref{fig:physics}(b) ($\left|\mathrm{Re}\left\{E_y\right\}\right|$ Appendix~\ref{Sec:Grav_Anal}), for offset (top panels) and facing (bottom panels) illuminations, with $r_\mathrm{s}=3.15\lambda_0=0.63$~mm (horizon radius), and $W_0=6.5\lambda_0$ for (a) and $W_0=8.5\lambda_0$ for (b).}
  \label{fig:Horizon}
\end{figure}

While the paper has focused onto a uniform-direction and constant-proper acceleration profile until this point, the reported accelerated-modulation space-time metamaterials may assume a virtually unlimited diversity of metrics, and potentially feature equally diverse opportunities for light manipulation. To illustrate this, Fig.~\ref{fig:Horizon} presents metamaterials that mimic Schwarzschild holes~\cite{schwarzschild1916uber}, based on an average refractive index profile that follows the corresponding metric~\cite{leonhardt1999optics}. The black hole, shown in Fig.~\ref{fig:Horizon}, displays the well-know light attraction and absorption effects, while the white hole, whose cosmological existence is purely theoretical but which may actually be implemented by an accelerated-modulation metamaterial, displays the complementary light deviation and repelling effects (see Appendix~\ref{Sec:Grav_Anal}).

\section{Conclusion}
This paper has initiated the field of accelerated-modulation space-time metamaterials. The capability of such materials to curve space-time of light, as gravitation does, augurs vast opportunities for scientific and technological development with multiple applications in electronics, electromagnetics and optics.
\appendix
\section{General Relativity Tools}\label{Sec:GR_Tools}

The key results in the paper are obtained using general relativistic electrodynamics, specifically the Maxwell-Cartan equations, the Rindler transformations of the space-time variables and the general tensorial coordinate transformation formulas.

The Maxwell-Cartan equations are generalizations of the Maxwell's equations for curved space-time. They 
read~\cite{cartan1924varietes,carroll2019spacetime}
\begin{subequations}\label{eq:Maxwell_Cartan}
\begin{equation}
	\partial_{(\lambda} F_{\mu \nu )}=
        \partial_\lambda F_{\mu\nu}+
        \partial_\nu F_{\lambda\mu}+
        \partial_\mu F_{\nu\lambda}=0
	\end{equation}
and
	\begin{equation}\label{eq:Maxwell_Cartan_G} 
	\partial_\mu G^{\mu \nu}=0,
	\end{equation}
\end{subequations}
where $F_{\mu \nu}$ is the Faraday tensor and $G^{\mu \nu}$ is its dual~\cite{van2012relativity}, which are defined as
\begin{subequations}\label{eq:Def_F_G}
\begin{equation}\label{eq:Def_F}
	F_{\mu \nu}=
	\begin{bmatrix}
		0 & E_x & E_y & E_z \\
		-E_x & 0 & -cB_z & cB_y\\
		-E_y & cB_z & 0 & -cB_x\\
		-E_z & -cB_y & cB_x & 0
	\end{bmatrix}
\end{equation}
and
\begin{equation}\label{eq:Def_G}
	G^{\mu \nu}=
	\begin{bmatrix}
		0 & -cD_x & -cD_y & -cD_z\\
		cD_x & 0 & H_z & -H_y\\
		cD_y & -H_z & 0 & H_x\\
		cD_z & H_y & -H_x & 0
	\end{bmatrix}.
\end{equation}
\end{subequations}
Note that the Einstein summation convention, whereby repeated indices imply summation over corresponding indices, are assumed everywhere in the paper, as for instance in~\eqref{eq:Maxwell_Cartan_G}, where $\partial_\mu G^{\mu \nu}=\sum_{\mu=0}^{3}\partial_\mu G^{\mu \nu}$.

The Rindler transformations are the coordinate transformations between the laboratory frame and the moving frame for the case of constant proper acceleration~\cite{rindler1960hyperbolic}. We shall use here the Kottler-M{\o}ller~\cite{moller1972theory} version of the Rindler transformations~\cite{rindler1960hyperbolic} to accommodate nonzero constant initial velocities. The Kottler-M{\o}ller transformations, assuming propagation in the $z$-direction, are given by the relations
\begin{subequations}\label{eq:Rinder_Transformations}
\begin{equation}\label{eq:Rindler_T}
	ct=\frac{c^2}{a'}\sqrt{g_{00}}\sinh(\xi+\xi_0)-\frac{c^2}{a'}\sinh(\xi_0),
\end{equation}
\begin{equation}
	x=x',
\end{equation}
\begin{equation}
	~y=y'
\end{equation}
and
\begin{equation}\label{eq:Rindler_Z}
	z=\frac{c^2}{a'}\sqrt{g_{00}}\cosh(\xi+\xi_0)-\frac{c^2}{a'}\cosh(\xi_0),
\end{equation}
where
\begin{equation}
	\xi=a't'/c
\end{equation}
and
\begin{equation}
	\xi_0=\sinh^{-1}(\beta_0\gamma_0),
\end{equation}
\end{subequations}
with $a'$ being the (constant) proper acceleration, $\beta_0$ and $\gamma_0$ being the initial relative velocity and Lorentz factor, respectively, and where
\begin{subequations}\label{eq:Rinder_Metric}
\begin{equation}\label{eq:Rinder_Metric_g}
	g_{00}=\left( 1+(a'z'/c^2)\right) ^2,
\end{equation}
\end{subequations}
is the $00$-term of the Rindler metric, which is simply $g_{\mu'\nu'}=\mathrm{diag}(g_{00},1,1,1)$. The corresponding ($K$-frame) instantaneous normalized velocity may be calculated by dividing~\eqref{eq:Rindler_Z} and~\eqref{eq:Rindler_T} by $dt'$, calculating the related derivatives and taking the ratio of the resulting expressions, which yields
\begin{equation}\label{eq:vel_K}
	\beta=\frac{dz}{d(ct)}=\frac{dz/d(ct')}{d(ct)/d(ct')}=\tanh(\xi+\xi_0),
\end{equation}
while the ($K$-frame) acceleration may be derived by taking the time derivative of this equation, which gives
\begin{equation}\label{eq:acc_K}
	a=a'\left(\cosh(\xi+\xi_0)\right)^{-3}.
\end{equation}

The inverse of~\eqref{eq:Rindler_T} and~\eqref{eq:Rindler_Z} may be found as follows. First, we isolate $\sinh(\xi+\xi_0)$ in~\eqref{eq:Rindler_T} and $\cosh(\xi+\xi_0)$ in~\eqref{eq:Rindler_Z}. Then, we obtain the relation $t'=t'(t)$ by successively taking the ratio of the resulting expressions and the inverse of the new expression, while we obtain the relation $z'=z'(z)$ by squaring the resulting expressions, taking the difference of the new expressions and taking the square root of the result. The final result is
\begin{subequations}\label{eq:Rinder_Inverse_Transformations}
\begin{equation}\label{eq:Rindler_T'}
	ct'=\frac{c^2}{a'}\tanh^{-1}\left(\frac{ct+\frac{c^2}{a'}\sinh(\xi_0)}{z+\frac{c^2}{a'}\cosh(\xi_0)}\right)-c\xi_0
\end{equation}
and
\begin{equation}\label{eq:Rindler_Z'}
	z'=\sqrt{(z+\frac{c^2}{a'}\cosh(\xi_0))^2-(ct+\frac{c^2}{a'}\sinh(\xi_0))^2}-\frac{c^2}{a'}.
\end{equation}
\end{subequations}

Finally, the tensorial coordinate transformations of the fields are given by the covariant and contravariant relations~\cite{tu2017differential}:
\begin{subequations}\label{eq:Covariant_Transformations}
\begin{equation}\label{eq:CT_a}
	F_{\mu' \nu'}=\frac{\partial x^\rho}{\partial x^{\mu'}}\frac{\partial x^\sigma}{\partial x^{\nu'}}F_{\rho \sigma}
\end{equation}
and
\begin{equation}
	\left|\text{det}\left(\frac{\partial x^{\mu'}}{\partial x^{\nu'}} \right) \right|G^{\mu' \nu'}=\frac{\partial x^{\mu'}}{\partial x^\rho}\frac{\partial x^{\nu'}}{\partial x^\sigma}G^{\rho \sigma}.
\end{equation}
\end{subequations}
For a rank-one tensor, such as the wave tensor ($k_\mu$), these relations reduce to
\begin{subequations}\label{eq:Rank1_Cov_Trans}
\begin{equation}\label{eq:RCT_a}
	k_{\mu'}=\frac{\partial x^\rho}{\partial x^{\mu'}}k_\rho
\end{equation}
and
\begin{equation}\label{eq:RCT_b}
	k^{\mu'}=\frac{\partial x^{\mu'}}{\partial x^\rho}k^\rho.
\end{equation}
\end{subequations}

\section{Accelerated-Matter Media}\label{Sec:Mat}
\subsection{Dispersion Relation}\label{Sec:Mat_Disp}
In the $K'$-frame, matter is stationary and isotropic. However, $K'$ is noninertial, which requires a general relativistic treatment~\cite{carroll2019spacetime}. In particular, the dispersion relation must be written in its covariant form~\cite{van1973relativistic}, viz.,
\begin{equation}\label{eq:Dis_Cov}
    k_{\mu'}k^{\mu'}=0,
\end{equation}
where
\begin{equation}\label{eq:k^mu}
    k^{\mu'}=\left(\frac{\omega'}{c},\mathbf{k'}/n'\right)
\end{equation}
is the contravariant form of the wave tensor, with $\mathbf{k}'$ being the wave vector, and the covariant form of the wave tensor is found by contraction with the metric tensor, which yields
\begin{equation}\label{eq:k_mu}
    k_{\mu'}=g_{\mu'\nu'}k^{\nu'}=\left(-g_{00}\frac{\omega'}{c},\mathbf{k'}/n'\right).
\end{equation}

Inserting then~\eqref{eq:k^mu} and~\eqref{eq:k_mu} into~\eqref{eq:Dis_Cov} yields the $K'$-frame dispersion relation
\begin{equation}\label{eq:Dips_Mat_K'_s}
    \frac{k'^2_z}{n'^2}+\frac{k'^2_x}{n'^2}=g_{00}\left(\frac{\omega'}{c}\right)^2.
\end{equation}
Finally, inverse-transforming~\eqref{eq:Dips_Mat_K'_s} with~\eqref{eq:Rindler_T} to~\eqref{eq:Rindler_Z} according to~\eqref{eq:Rank1_Cov_Trans} leads to the $K$-frame dispersion relation
\begin{subequations}\label{eq:Dispersion_Matter_Isotorpic}
\begin{equation}
	 \frac{(k_z-\rchi \omega/c)^2}{\alpha^2 n'^2}+\frac{k_x^2}{\alpha n'^2}=\left(\frac{\omega}{c}\right)^2,
\end{equation}
where
\begin{equation}
	 \alpha=(1-\beta^2)/(1-n'^2\beta^2)
\end{equation}
and
\begin{equation}
	 \rchi=\beta(1-n'^2)/(1-n'^2 \beta^2),
\end{equation}
\end{subequations}
with $n'$ being the refractive index of the medium measured in the $K'$-frame.

\subsection{Fields and Deflection Angles}\label{Sec:Mat_Fields}
\begin{figure}[h!] 
   \centering
  \includegraphics[width=0.3\textwidth]{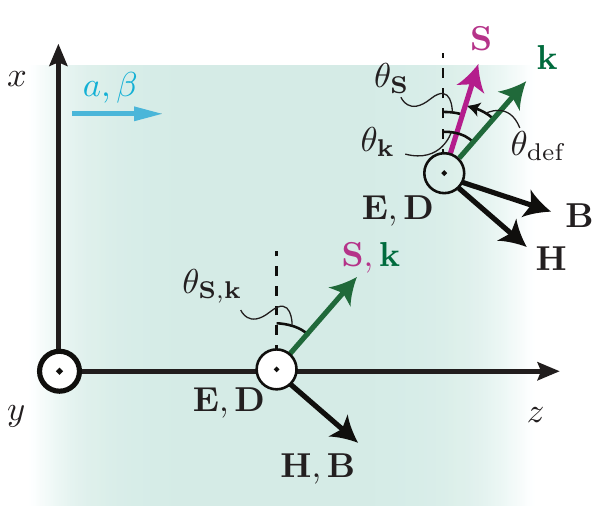}
  \caption{Incident and propagating fields configuration.}
  \label{fig:Incidence}
\end{figure}

Let us now consider an incident plane wave in vacuum, which will be later used as an ansatz to calculate the fields of interest. Under the assumption of s polarization, depicted in Fig.~\ref{fig:Incidence}, such a wave satisfies the relations $E_x=E_z=B_y=D_x=D_z=H_y=0$, and its Faraday tensor and dual Faraday tensor are then found, upon inserting these relations into Eqs.~\eqref{eq:Def_F_G}, as
\begin{subequations}\label{eq:F_G_vacuum_K}
\begin{equation}\label{eq:F_vacuum_K}
	F_{\mu \nu}=
	\begin{bmatrix}
		0 & 0 & 1 & 0 \\
		0 & 0 & -\cos \theta & 0\\
		-1 & \cos \theta & 0 & \sin \theta\\
		0 & 0 & -\sin \theta & 0
	\end{bmatrix}E_0 f\left( \zeta \right)
    \end{equation}
    and
    \begin{equation}\label{eq:G_vacuum_K}
	G^{\mu \nu}=
	\begin{bmatrix}
		0 & 0 & -\eta_0^{-1} & 0 \\
		0 & 0 & \frac{\cos\theta}{\eta_0} & 0\\
		\eta_0^{-1} & -\frac{\cos\theta}{\eta_0} & 0 & \frac{\sin\theta}{\eta_0}\\
		0 & 0 & -\frac{\sin\theta}{\eta_0} & 0
	\end{bmatrix}E_0 f\left( \zeta \right),
    \end{equation}
after defining
    \begin{equation}
        \zeta=z\sin{\theta}+x\cos{\theta}-ct,
    \end{equation}
\end{subequations}
where $f(\cdot)$ and $\theta$ are the incident (initial) wave profile and angle, respectively.
The fields in the $K$-frame may be derived by solving the wave equation in the $K'$-frame and inverse-Rindler transforming the results to the $K$-frame, which results in~\cite{tanaka1978relativistic}
\begin{subequations}\label{eq:F_G_matter_K}
\begin{equation}\label{eq:F_matter_K}
\begin{split}
    &F_{\mu \nu}=\gamma \frac{P_n}{n'}E_0 f(\zeta)\\
	&\begin{bmatrix}
		0 & 0 & 1 + n'\beta\sin\theta'_n & 0 \\
		0 & 0 & \frac{n'\cos{\theta'_n}}{\gamma} & 0\\
		-(1 + n'\beta \sin \theta'_n) & -\frac{n' \cos{\theta'_n}}{\gamma} & 0 & (n'\sin \theta'_n + \beta)\\
		0 & 0 & -(n'\sin\theta'_n+\beta) & 0
	\end{bmatrix}
\end{split}
\end{equation}
and
\begin{equation}\label{eq:G_matter_K}
\begin{split}
    &G^{\mu \nu}=\gamma P_n E_0 f(\zeta)\\
	&\begin{bmatrix}
		0 & 0 & -\frac{n'+\beta\sin\theta'_n}{\eta_0\mu'} & 0 \\
		0 & 0 &  \frac{\cos\theta'_n}{\gamma\eta_0 \mu'} &  0\\
		\frac{n'+\beta\sin\theta'_n}{\eta_0 \mu'} & -\frac{\cos\theta'_n}{\gamma\eta_0\mu'} & 0 &  -\frac{\sin\theta'_n+n'\beta}{\eta_0\mu'}\\
	0 & 0 & \frac{\sin\theta'_n+n'\beta}{\eta_0\mu'} & 0
	\end{bmatrix},
\end{split}
\end{equation}
where 
    \begin{equation}\label{eq:P_n_Def}
      P_n=\cosh(\xi/n'+\xi_0)-\sin(\theta)\sinh(\xi/n'+\xi_0),
    \end{equation}
    \begin{equation}\label{eq:theta'_n_Def}
      \cos\theta'_n=\frac{\cos\theta}{P_n}
    \end{equation}
and
    \begin{equation}
     \gamma=1/\sqrt{1-\beta^2}=\cosh(\xi+\xi_0),
    \end{equation}
\end{subequations}
with $\beta$ and $n'=\sqrt{\epsilon'\mu'}$ being the instantaneous velocity and the stationary refractive index of the medium.

The Poynting vector is given by
\begin{equation}\label{eq:Poynting_Gen}
    \mathbf{S}=\mathbf{E} \times \mathbf{H}=-E_y H_x \hat{z}+E_y H_z \hat{x},
\end{equation}
and its angle with respect to the $x$-axis can be written as
\begin{equation} \label{eq:Angle_Poynting_Mat_an}
    \theta_\mathbf{S}
    =-\tan^{-1}\left(\frac{E_yH_x}{E_yH_z}\right)
    =-\tan^{-1}\left(\frac{H_x}{H_z}\right), 
\end{equation}
where $H_z$ and $H_x$ are found in~\eqref{eq:G_matter_K} with the help of~\eqref{eq:Def_G}, which results into
\begin{equation}\label{eq:Angle_Poynting_Mat_is}
    \theta_\mathbf{S}=\tan^{-1}\left( \gamma \frac{\sin \theta_n' + n' \beta}{\cos \theta_n'} \right).
\end{equation}
where $\theta'_n$ is given in~\eqref{eq:theta'_n_Def}. 

The deflection angle is the angle the Poynting vector makes with the phase angle ($\theta_\mathbf{k}=\tan^{-1}(k_z/k_x)$). Therefore, it can be represented as $\theta_\mathrm{defl}=\theta_\mathbf{S}-\theta_\mathbf{k}$, and explicitly written
\begin{equation}\label{eq:Mat_Defl_angle}
    \theta_\mathrm{defl}=\tan^{-1}{\left(\frac{\gamma(Q_n+n'\beta P_n)-\sin\theta}{\cos\theta+\gamma (Q_n+n'\beta P_n)\tan\theta}\right)},
\end{equation}
where
\begin{equation}\label{eq:Q_n_Def}
    Q_n=\sin\theta\cosh(\xi_n'+\xi_0)-\sinh(\xi_n'+\xi_0),
\end{equation}
and $P_n$ is given in~\eqref{eq:P_n_Def}.

\section{Accelerated-Modulation Media}\label{Sec:Mod}

\subsection{Dispersion Relation}\label{Sec:Mod_Disp}

Under the assumption of (moving-\emph{modulation}) layer isotropy in the $K$-frame, the $K$-frame constitutive relations read 
\begin{subequations}\label{eq:Constitutive_K_Loc}
 	\begin{equation}
	\mathbf{B}_{1,2}=\mu_{1,2} \mathbf{H}_{1,2}
	\end{equation}
    and
 	\begin{equation}
	\mathbf{D}_{1,2}=\epsilon_{1,2} \mathbf{E}_{1,2}.
\end{equation}
\end{subequations}
Rindler transforming~\eqref{eq:Constitutive_K_Loc} to the $K'$-frame using~\eqref{eq:Covariant_Transformations} with~\eqref{eq:Rinder_Transformations} yields, for each of the two layers, the bianisotropic constitutive relations

\begin{subequations}\label{eq:Constitutive_K'_bilayer}
    \begin{equation}\label{eq:Constitutive_K'_bilayer_B} \sqrt{g_{00}}\mathbf{B}'_{1,2}=\mu_0\ov{\ov{\mu}}'_{1,2}\mathbf{H'}_{1,2}+\ov{\ov{\rchi}}'_{1,2}\mathbf{E}'_{1,2}
    \end{equation}
    and
    \begin{equation}\label{eq:Constitutive_K'_bilayer_D} 
    \sqrt{g_{00}}\mathbf{D}'_{1,2}=\epsilon_0\ov{\ov{\epsilon}}'_{1,2}\mathbf{E}'_{1,2}-\ov{\ov{\rchi}}'_{1,2}\mathbf{H}'_{1,2},
    \end{equation}
\end{subequations}
with
\begin{subequations}
\begin{equation}	
    \ov{\ov{\mu}}'_{1,2}=
		\begin{bmatrix}
		\alpha'_{1,2} \mu'_{1,2} & 0 & 0\\
		0 & \alpha'_{1,2} \mu'_{1,2} & 0\\
		0 & 0 & \mu'_{1,2}
	\end{bmatrix},
\end{equation}
\begin{equation} 	
        \ov{\ov{\epsilon}}'_{1,2}=
	\begin{bmatrix}
		\alpha'_{1,2} \epsilon'_{1,2} & 0 & 0\\
		0 & \alpha'_{1,2} \epsilon'_{1,2} & 0\\
		0 & 0 & \epsilon'_{1,2}
	\end{bmatrix}
  \end{equation}
        and
\begin{equation}
         \ov{\ov{\rchi}}'_{1,2}=
	\begin{bmatrix}
		0 & \rchi'_{1,2}/c & 0\\
		-\rchi'_{1,2}/c & 0 & 0\\
		0 & 0 & 0
	\end{bmatrix},
\end{equation}
where
\begin{equation}
	 \alpha'_{1,2}=(1-\beta^2)/(1-\epsilon'_{1,2}\mu'_{1,2}\beta^2)
\end{equation}
and
\begin{equation}
	 \rchi'_{1,2}=-\beta(1-\epsilon'_{1,2}\mu'_{1,2})/(1-\epsilon'_{1,2}\mu'_{1,2} \beta^2).
\end{equation}
\end{subequations}
Averaging these relations via the continuity of the field $\mathbf{E}'_\|$, $\mathbf{H}'_\|$, $D'_\perp$ and $B'_\perp$ at the layer interfaces yields the homogenized constitutive relations
\begin{subequations}\label{eq:Constitutive_K'_avg}
 	\begin{equation}\label{eq:Constitutive_K'_avg_B}
\begin{split}
    \sqrt{g_{00}}\begin{bmatrix}
		 B'_x \\
		 B'_y \\
		 B'_z
	 \end{bmatrix}=&
		\mu_0\begin{bmatrix}
		\ov\mu'_\| & 0 & 0\\
		0 & \ov\mu'_\| & 0\\
		0 & 0 & \ov\mu'_\perp
	\end{bmatrix}
	\begin{bmatrix}
		 H'_x \\
		 H'_y \\
		 H'_z
	\end{bmatrix}
	\\&+\begin{bmatrix}
		0 & \ov{\rchi'}/c & 0\\
		-\ov{\rchi'}/c & 0 & 0\\
		0 & 0 & 0
	\end{bmatrix}
	\begin{bmatrix}
		 E'_x \\
		 E'_y \\
		 E'_z
	\end{bmatrix}
\end{split}
	\end{equation}
	and
 	\begin{equation}\label{eq:Constitutive_K'_avg_D}
\begin{split}
    \sqrt{g_{00}}\begin{bmatrix}
		 D'_x \\
		 D'_y \\
		 D'_z
	 \end{bmatrix}=&
	\epsilon_0	\begin{bmatrix}
		\ov\epsilon'_\| & 0 & 0\\
		0 & \ov\epsilon'_\| & 0\\
		0 & 0 & \ov\epsilon'_\perp
	\end{bmatrix}
	\begin{bmatrix}
		 E'_x \\
		 E'_y \\
		 E'_z
	\end{bmatrix}
		\\&+\begin{bmatrix}
		0 & -\ov{\rchi'}/c & 0\\
		\ov{\rchi'}/c & 0 & 0\\
		0 & 0 & 0
	\end{bmatrix}
	\begin{bmatrix}
		 H'_x \\
		 H'_y \\
		 H'_z
	\end{bmatrix},
\end{split}
\end{equation}
\end{subequations}
where
\begin{subequations}\label{eq:const_av_prime}
\begin{equation}\label{eq:eps_p_par}
	\ov\epsilon_\|'=(\alpha'_1\epsilon_1+\alpha'_2\epsilon_2)/2,
\end{equation} 
\begin{equation}\label{eq:eps_p_perp}
	\ov\epsilon_\perp'=2(\epsilon_1^{-1}+\epsilon_2^{-1})^{-1},
\end{equation}
\begin{equation}\label{eq:mu_p_par}
	\ov\mu_\|'=(\alpha'_1\mu_1+\alpha'_2\mu_2)/2,
\end{equation} 
\begin{equation}\label{eq:eps_p_perp}
	\ov\mu_\perp'=2(\mu_1^{-1}+\mu_2^{-1})^{-1}
\end{equation}
and
\begin{equation}\label{eq:chi_p}
	\overline{\rchi'}=(\rchi'_1+\rchi'_2)/2.
\end{equation} 
\end{subequations}

Following its homogenization, the bilayer-unit-cell moving-modulation medium at hand has become a bulk medium, whose general dispersion relation is well-known and is found by inserting the bulk constitutive relations into the wave equation and taking the Fourier transform of the resulting expression~\cite{kong1990electromagnetic}. This relation reads
\begin{equation}\label{eq:Dispersion_Bianisotropic}
	\left| \omega^2 \overline{\ov{\epsilon}}+[\overline{\overline{k}}+\omega \overline{\overline{\xi}}] \cdot \overline{\ov{\mu}}^{-1} \cdot [\overline{\overline{k}}-\omega \overline{\overline{\zeta}}]\right|=0.
\end{equation}
%
%
Inverse-Rindler transforming~\eqref{eq:Constitutive_K'_avg_B} and ~\eqref{eq:Constitutive_K'_avg_D} to the $K$-frame using~\eqref{eq:Covariant_Transformations} with~\eqref{eq:Rinder_Transformations} and inserting the resulting equation into~\eqref{eq:Dispersion_Bianisotropic} yields

 \begin{equation}\label{eq:Disp_Rel_K_Mod}
	 \frac{\left( k_z-\overline{\rchi} \omega/ c\right)^2}{\ov\epsilon_\|\,\ov\mu_\|}+\frac{k_x^2}{\ov\epsilon_\| \ov\mu_\perp}=\left(\frac{\omega}{c}\right)^2,
 \end{equation}
where
 \begin{subequations}\label{eq:param_av_all}
 \begin{equation}\label{eq:eps_tan_avg_mod}
     \ov\epsilon_\|=\ov\epsilon_\|'\frac{(1-\beta^2)}{(1-\beta\overline{\rchi'})^2-\ov\epsilon_\|'\,\ov\mu_\|'c^2 \beta^2)},
 \end{equation}
  \begin{equation}\label{eq:eps_norm_avg_mod}
\ov\epsilon_\perp=2(\epsilon_1^{-1}+\epsilon_2^{-1})^{-1},
 \end{equation}
  \begin{equation}\label{eq:mu_tan_avg_mod}
     \ov\mu_\|=\ov\mu_\|'\frac{1-\beta^2}{(1-\overline{\rchi'}\beta)^2-\ov\epsilon_\|'\,\ov\mu_\|'c^2 \beta^2},
 \end{equation}
 \begin{equation}\label{eq:mu_norm_avg_mod}
	\ov\mu_\perp=2(\mu_1^{-1}+\mu_2^{-1})^{-1}
 \end{equation}
 and
\begin{equation}\label{eq:chi_avg_mod}
     \overline{\rchi}=\frac{((\overline{\rchi'}-\beta)(1-\overline{\rchi'} \beta)+\epsilon_\|'\,\mu_\|'c^2 \beta)}{((1-\overline{\rchi'} \beta)^2-\epsilon_\|'\,\mu_\|'c^2\beta^2)}.
 \end{equation}
 \end{subequations}
We can simplify~\eqref{eq:param_av_all} by substituting~\eqref{eq:const_av_prime}, which leads to
\begin{subequations}
\begin{equation}
    \ov\epsilon_\|=\frac{\Sigma_\epsilon-\beta^2\epsilon_1\epsilon_2\Sigma_\mu}{1-\beta^2\Sigma_\epsilon\Sigma_\mu},
\end{equation}
\begin{equation}
    \epsilon_\perp=\frac{2\epsilon_1\epsilon_2}{\Sigma_\epsilon},
\end{equation}
\begin{equation}
    \ov\mu_\|=\frac{\Sigma_\mu-\beta^2\mu_1\mu_2\Sigma_\epsilon}{1-\beta^2\Sigma_\epsilon\Sigma_\mu},
\end{equation}
\begin{equation}
    \mu_\perp=\frac{2\mu_1\mu_2}{\Sigma_\mu}
\end{equation}
and
\begin{equation}\label{eq:chibmod}
    \ov{\rchi}=\beta\frac{\Delta_\epsilon\Delta_\mu}{1-\beta^2\Sigma_\epsilon\Sigma_\mu},
\end{equation}
\end{subequations}
where $\Sigma_\epsilon=(\epsilon_1+\epsilon_2)/2$, $\Delta_{\epsilon}=(\epsilon_1-\epsilon_2)/2$ and similarly for $\Sigma_\mu$, $\Delta_\mu$.

 \subsection{Deflection Angles}\label{Sec:Mod_Defl}
 
 The fields are solutions to the equations~\eqref{eq:Maxwell_Cartan}, which, upon substitution of~\eqref{eq:Def_F_G}, take the explicit form
\begin{subequations}\label{eq:Maxwell_Dyadic}
\begin{equation}
    \varepsilon^{ijk} \partial_j E_k=-\frac{\partial}{\partial t} B^i
\end{equation}
and
\begin{equation}
    \varepsilon^{ijk}\partial_j H_k=\frac{\partial}{\partial t} D^i,
\end{equation}
\end{subequations}
where $\varepsilon^{ijk}$ (not to be confused with the permittivity tensor, $\epsilon^{ij}$) is the Levi-Civita pseudo-tensor~\cite{berkshire1979introduction} and where the Latin indices run from 1 to 3 (instead of 0 to 3 for the Greek indices in what precedes).

On the other hand, the constitutive relations may be generally expressed in the bianisotropic form~\cite{kong1990electromagnetic}
\begin{subequations}\label{eq:Constitutive_Dyadic}
\begin{equation}\label{eq:Constitutive_Dyadic_D} 
\tilde{D}^i=\epsilon^{ij}\tilde{E}_j+\varepsilon^{ijk}\rchi_j\tilde{H}_k
\end{equation}
and
\begin{equation}\label{eq:Constitutive_Dyadic_B}
\tilde{B}^i=\mu^{ij}\tilde{H}_j+\varepsilon^{jik}\rchi_j\tilde{E}_k,
\end{equation}
\end{subequations}
where $\epsilon^{ij}$ and $\mu^{ij}$ are the diagonal tensors
\begin{subequations}\label{eq:enc_par_perp}
\begin{equation}
   \epsilon^{ij}=\epsilon_0\, \mathrm{diag}(\epsilon_\|,\epsilon_\|,\epsilon_\perp)
\end{equation}
and
\begin{equation}
    \mu^{ij}=\mu_0\, \mathrm{diag}(\mu_\|,\mu_\|,\mu_\perp),
\end{equation}
and where
\begin{equation}
    \rchi_j= (0,0,\ov{\rchi}/c),
\end{equation}
\end{subequations}
with the different tensor components being given by~\eqref{eq:eps_tan_avg_mod} to \eqref{eq:chi_avg_mod}.

In order to combine~\eqref{eq:Maxwell_Dyadic} and~\eqref{eq:Constitutive_Dyadic}, we write~\eqref{eq:Maxwell_Dyadic} in the Fourier domain, viz., 
\begin{subequations}\label{eq:Maxwell_Dyadic_k}
\begin{equation}
    \varepsilon^{ijk} k_j \Tilde{E}_k=\omega \Tilde{B}^i
\end{equation}
and
\begin{equation}
    \varepsilon^{ijk} k_j \Tilde{H}_k=-\omega \Tilde{D}^i.
\end{equation}
\end{subequations}
Substituting these relations into~\eqref{eq:Constitutive_Dyadic}, and separating the $\tilde{E}_j$ and $\tilde{H}_j$ in each of the resulting equations yields
\begin{subequations}\label{eq:EH_Dyadic}
\begin{equation}\label{eq:E_Dyadic_gen}
     \epsilon^{ij}\tilde{E}_j=-\frac{1}{\omega}\left(\varepsilon^{ijk} k_j+\omega \varepsilon^{ijk}\rchi_j\right)\tilde{H}_k
\end{equation}
and
\begin{equation}\label{eq:H_Dyadic_gen}
    \mu^{ij}\tilde{H}_j=\frac{1}{\omega}\left(\varepsilon^{ijk}k_j-\omega \varepsilon^{jik}\rchi_j\right)\tilde{E}_k.
\end{equation}
\end{subequations}

Inserting~\eqref{eq:eps_tan_avg_mod} to~\eqref{eq:chi_avg_mod} into~\eqref{eq:enc_par_perp}, and substituting the resulting expressions into~\eqref{eq:EH_Dyadic} yields
\begin{subequations}\label{eq:Fields_K_Modulation}
\begin{equation}\label{eq:Ey_Fields_K_Modulation}
     \tilde{E}_y=-\frac{k_z-\omega \overline{\rchi}/c}{\omega \epsilon_\|}\tilde{H}_x+\frac{k_x}{\omega \epsilon_\|}\tilde{H}_z,
\end{equation}
\begin{equation}\label{eq:Hx_Fields_K_Modulation}
    \tilde{H}_x=-\frac{k_z-\omega \overline{\rchi}/c}{\omega \mu_\|}\tilde{E}_y
\end{equation}
and
\begin{equation}\label{eq:Hz_Fields_K_Modulation}
    \tilde{H}_z=\frac{k_x}{\omega \mu_\perp}\tilde{E}_y.
\end{equation}
\end{subequations}
The $H_x$ and $H_z$ in the direct domain are then found as the inverse-Fourier transform of~\eqref{eq:Hx_Fields_K_Modulation} and\eqref{eq:Hz_Fields_K_Modulation}, respectively.
 
The spectral Poynting vector is given by
\begin{equation}\label{eq:Poynting_dyadic}
    \mathbf{\tilde{S}}=\mathbf{\tilde{E}} \times \mathbf{\tilde{H}}=-\tilde{E_y} \tilde{H_x} \hat{z}+\tilde{E_y} \tilde{H_z} \hat{x},
\end{equation}
and its angle with respect to the $x$-axis can then be written as
\begin{equation}\label{eq:Poynting_Angles}
    \theta_{\mathbf{S}}
    =-\tan^{-1}\left(\frac{\tilde{E_y}\tilde{H}_x}{\tilde{E}_y\tilde{H}_z}\right)=-\tan^{-1}\left(\frac{\tilde{H}_x}{\tilde{H}_z}\right).
\end{equation}
Substituting~\eqref{eq:Hx_Fields_K_Modulation} and~\eqref{eq:Hz_Fields_K_Modulation} into this relation yields
\begin{equation}\label{eq:Poynting_Angles}
    \theta_{\mathbf{S}}
    =\tan^{-1}\left({\frac{\ov\mu_\perp}{\ov\mu_\|}\frac{k_z-\omega\overline{\rchi}/c}{k_x}} \right),
\end{equation}
which, using $k_x=k_0\cos\theta$ and $k_z=k_0\sin\theta$, may also be written as
\begin{equation}\label{eq:Poynting_Angles_theta}
    \theta_{\mathbf{S}}
    =\tan^{-1}\left({\frac{\ov\mu_\perp}{\ov\mu_\|}\frac{\sin\theta-\ov{\rchi}}{\cos\theta}} \right).
\end{equation}
The deflection angle, defined as the difference between the phase (launching) direction and the power direction, is then obtained as $\theta_{\mathrm{defl}}=\theta_{\mathbf{S}}-\theta_{\mathbf{k}}$, where $\theta_{\mathbf{k}}=\arctan(k_z/k_x)$ is the phase angle. Using the identity $\tan^{-1}(x)-\tan^{-1}(y)=\tan^{-1}((x-y)/(1+xy))$, this angle may be expressed as
\begin{equation}\label{eq:Deflection_Ang_sup}
    \theta_{\mathrm{defl}}=\tan^{-1}\left(\frac{(\mu_\perp-\mu_\|) k_x k_z-\mu_\perp k_x \omega \overline{\rchi}/c}{\mu_\perp k_x^2+\mu_\perp k_z^2-k_z \omega \overline{\rchi}/c}\right).
\end{equation}

\subsection{Fields for Gaussian Beam Excitation}\label{Sec:Mod_Fields}

We consider the propagation of an s polarized plane-wave Gaussian beam that is launched in the $x$ direction. The corresponding field in the plane $x=0$ reads
\begin{equation}\label{eq:Gauss_Initial_dir}
    E_y(z,x=0)=E_0\exp\left(-\frac{z^2}{W_0^2}\right),
\end{equation}
where $E_0$ and $W_0$ are the amplitude and the waist of the beam. Taking the $z$-spatial Fourier transform of~\eqref{eq:Gauss_Initial_dir} yields
\begin{equation}\label{eq:Gauss_Initial_inv}
\begin{split}
    \tilde{E}_y(k_z; x=0)&=\frac{1}{2\pi}\int_{-\infty}^{+\infty} E_0\exp\left(-\frac{z^2}{W_0^2}\right) \text{e}^{-\mathrm{i}k_zz}dz\\&=\frac{W_0E_0}{2\sqrt{\pi}}\exp\left(-\frac{k_z^2W_0^2}{4}\right).
\end{split}
\end{equation}
The spectral field in any plane $x$ may be obtained by multiplying~\eqref{eq:Gauss_Initial_inv} with the propagator of the problem~\cite{novotny2012principles}, $\text{e}^{ik_xx}$, namely
\begin{equation}\label{eq:propagator_imp_inv}
    \tilde{E}_y(k_z, x)=\text{e}^{\mathrm{i}k_xx}\tilde{E}_y(k_z,x=0),
\end{equation}
where $k_x$ is related to $k_z$ by~\eqref{eq:Disp_Rel_K_Mod}. The field in~\eqref{eq:Gauss_Initial_inv} becomes then
\begin{equation}\label{eq:propagator_exp_inv}
    \tilde{E}_y(k_z, x)=\frac{W_0E_0}{2\sqrt{\pi}}\exp\left(-\frac{k_z^2W_0^2}{4}\right)\text{e}^{\mathrm{i}k_xx},
\end{equation}
to which we apply the inverse Fourier transform to provide the field at any point $z$ and $x$,
\begin{equation}\label{eq:propagator_exp_dir}
    E_y(z,x)=\frac{W_0E_0}{2\sqrt{\pi}}\int_{-\infty}^{+\infty}\exp\left(-\frac{k_z^2W_0^2}{4}\right)\text{e}^{\mathrm{i}(k_xx+k_zz)}dk_z.
\end{equation}
Given the assumed $x$-direction of propagation and the Gaussian nature of beam, $k_z\ll k_0$ ($k_0=\omega/c$), $k_x$ in~\eqref{eq:Disp_Rel_K_Mod} may be approximated as
\begin{equation}\label{eq:paraaxial_approx}
    k_x\approx\sqrt{\ov\epsilon_\|\ov\mu_\perp}k_0\left(1-\frac{(k_z-\ov{\rchi}k_0)^2}{2\ov\epsilon_\|\ov\mu_\|k_0^2}\right),
\end{equation}
which is a modulation-motion bianisotropic generalization of the paraxial approximation.
Inserting this expression into~\eqref{eq:propagator_exp_dir}, rearranging and factoring out the $k_z$-independent terms yields
\begin{equation}\label{eq:E_dir_rearranged}
\begin{split}
    E_y(z,x)=&\frac{E_0W_0}{2\sqrt{\pi}}\text{e}^{\mathrm{i}\sqrt{\ov\epsilon_\|\ov\mu_\perp}k_0\left(1-\tfrac{\ov{\chi}^2k_0}{2\ov\epsilon_\|\ov\mu_\|}\right)x}\\&
    \quad\int_{-\infty}^{+\infty}\exp\left(-k_z^2\left(\frac{W_0^2}{4}+\mathrm{i}\frac{\sqrt{\ov\epsilon_\|\ov\mu_\perp}}{\ov\epsilon_\|\ov\mu_\|}\frac{x}{2k_0}\right)\right)\\&
    \quad\quad\text{e}^{\mathrm{i}\frac{\sqrt{\ov\epsilon_\|\ov\mu_\perp}}{\ov\epsilon_\|\ov\mu_\|}\ov{\rchi}xk_z}\text{e}^{ik_zz}dk_z.
\end{split}
\end{equation}
Calculating the integral in~\eqref{eq:E_dir_rearranged} yields the final result,
\begin{equation}\label{eq:E_dir_final}
\begin{split}
E_y(z,x)=&E_0\frac{\text{e}^{\mathrm{i}\sqrt{\ov\epsilon_\|\ov\mu_\perp} k_0\left(1-\tfrac{\ov{\chi}^2k_0}{2\ov\epsilon_\|\ov\mu_\|}\right)x}}{1+\mathrm{i}\frac{\sqrt{\ov\epsilon_\|\ov\mu_\perp}}{\ov\epsilon_\|\ov\mu_\|}\frac{2x}{k_0W_0^2}}
    \\&\quad\exp\left(-\frac{\left(z+\frac{\sqrt{\ov\epsilon_\|\ov\mu_\perp}}{\ov\epsilon_\|\ov\mu_\|}\ov{\rchi}x\right)^2}{\left(1+\mathrm{i}\frac{\sqrt{\ov\epsilon_\|\ov\mu_\perp}}{\ov\epsilon_\|\ov\mu_\|}\frac{2x}{k_0W_0^2}\right)W_0^2}\right),
\end{split}
\end{equation}
which features an $x$-dependent $z$ lateral shift.

Note that in the absence of modulation ($\beta=0$ and two identical layers), the medium reduces to a simple \emph{isotropic stationary medium}, with~\eqref{eq:E_dir_final} correspondingly reducing to
\begin{equation}\label{eq:E_dir_Iso_reduc}
\begin{multlined}
    E_y(z,x)=E_0\frac{\text{e}^{\mathrm{i}kx}}{1+\mathrm{i}\frac{2x}{kW_0^2}}\exp\left(-\frac{z^2}{\left(1+\mathrm{i}\frac{2x}{kW_0^2}\right)W_0^2}\right),
\end{multlined}
\end{equation}
where $k=nk_0$, where $n$ is the refractive index of the medium~\cite{novotny2012principles}.

\subsection{Numerical Validation}\label{Sec:FDTD}
\begin{figure}[h!] 
   \centering
  \includegraphics[width=0.5\textwidth]{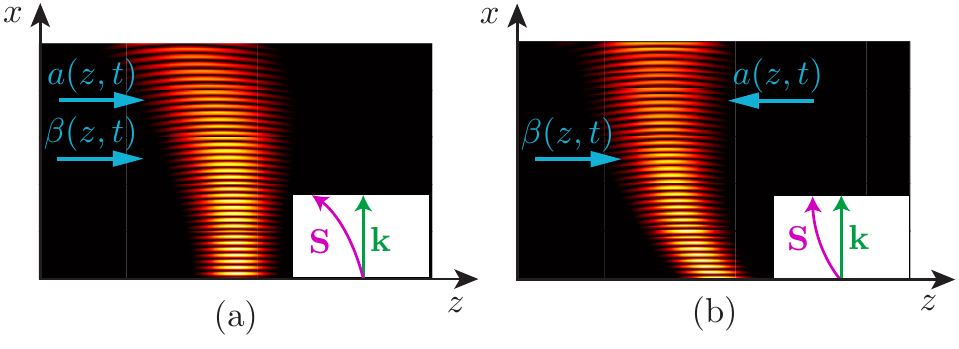}
  \caption{FDTD simulation corresponding to the analytical result in Fig.~3, with FDTD mesh size $\Delta z=\Delta x=(3/2)10^{-5}$~m, time step $\Delta t=(2/3)10^{-13}$~s, Mur's absorbing boundary conditions and a computational domain of area $417\Delta z\times 250\Delta x$ and duration $33000\Delta t$.}
  \label{fig:FDTD_Defl}
\end{figure}

Figure~\ref{fig:FDTD_Defl} shows the FDTD simulation corresponding to the analytical result in Fig.~3.

\begin{figure}[!h] 
    \centering
  \includegraphics[width=0.5\textwidth]{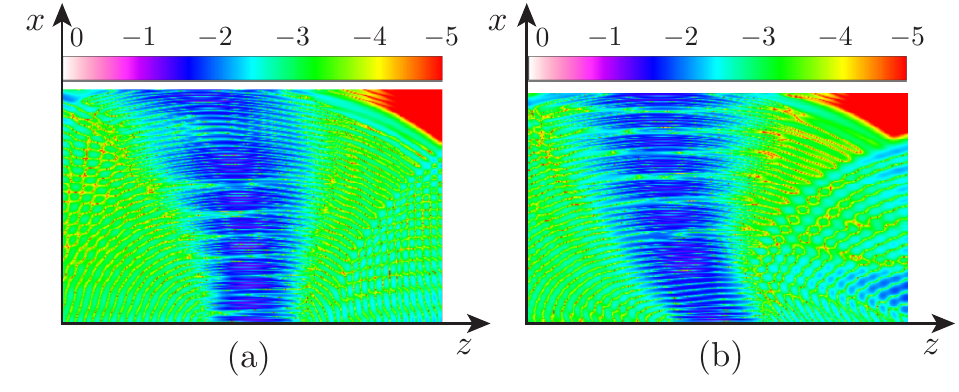}
  \caption{Difference $\log_{10}(\delta_{E_y})$, with range limited to the interval $[0,-5]$~dB, corresponding to $\max\left(\delta_{E_y}\right)=5\%$, between the analytical field plotted in Fig.~3 and the FDTD field plotted in Fig.~\ref{fig:FDTD_Defl} . 
  }
  \label{fig:Errors}
\end{figure}

Figure~\ref{fig:Errors} plots the logarithm of the difference between the analytical field plotted in Fig.~3 and the FDTD field plotted in Fig.~\ref{fig:FDTD_Defl}, with this difference being as defined as
\begin{equation}\label{eq:error}
    \delta_{E_y}=\frac{\left|E_y^{\mathrm{analytical}}-E_y^{\mathrm{FDTD}}\right|}{E_0}.
\end{equation}
We verified that the difference in Fig.~\ref{fig:Errors} decreases with increasing FDTD mesh density. This convergence of the FDTD simulation to the analytical result validates the latter.

\section{Modulation Parameters for Black and White Holes}\label{Sec:Grav_Anal}
The electromagnetic parameters corresponding to a given gravitational metric $g_{\mu\nu}$ are~\cite{plebanski1960electromagnetic}
\begin{subequations}\label{eq:Grav_Analog}
\begin{equation}\label{eq:Grav_Analog_eps_mu}
\epsilon^{ij}=\mu^{ij}=-\frac{\sqrt{-g}}{g_{00}} g^{ij}
\end{equation}
for the permittivity and permeability, and
\begin{equation}\label{eq:Grav_Analog_chi}
\rchi_j=\frac{g_{0j}}{g_{00}}
\end{equation}
\end{subequations}
for the chirality.

The metric of black or white holes is found  from the Schwarzschild space-time interval
\begin{subequations}\label{eq:sti}
\begin{equation}\label{eq:ds2_Schw}
    ds^2=-\left(1-\frac{2MG}{r}\right)(cdt)^2+\left(1-\frac{2MG}{r}\right)^{-1}dr^2+r^2d\Omega^2,
\end{equation}
where 
\begin{equation}\label{eq:BH_metric_param}
    d\Omega^2=d\theta^2 + \sin^2 \theta d\phi^2,
\end{equation}
\end{subequations}
as
\begin{subequations}\label{eq:BH_metric_main}
\begin{equation}
    g_{00}=-\left(1-\frac{2MG}{r}\right)
\end{equation}
and
\begin{equation}
    g_{11}=\left(1-\frac{2MG}{r}\right)^{-1},\quad
    g_{22}=r^2,\quad
    g_{33}=r^2\sin^2\theta.
\end{equation}
\end{subequations} 

The Schwarzschild metric~\eqref{eq:BH_metric_main} has two singularities, one at the origin, $r=0$, and one at the ``Schwarzschild radius'', $r=2MG$. The former is an essential singularity~\cite{penrose1965gravitational}, whereas the latter is an artifact of the coordinates system in~\eqref{eq:sti}~\footnote{The metric associated with~\eqref{eq:sti} is not conformally flat, i.e., its manifold cannot be mapped to a flat space via a conformal transformation map~\cite{tu2017differential}} and can therefore be removed with a proper change of coordinates~\cite{kruskal1960maximal}. An appropriate coordinate system to remove the spatial singularity would be one with the radial coordinate being transformed as~\cite{weinberg1972gravitation}
\begin{equation}\label{eq:Conf_Trans}
    r=\hat{r}\left(1+\frac{2MG}{4\hat{r}}\right),
\end{equation}
where $\hat{r}$ is the new radial coordinate. Inserting~\eqref{eq:Conf_Trans} into~\eqref{eq:ds2_Schw} leads to the new space-time interval
\begin{subequations}

\begin{equation}\label{eq:sti_conf}
    ds^2=-\left(\frac{1-\frac{2MG}{4\hat{r}}}{1+\frac{2MG}{4\hat{r}}}\right)^2(cdt)^2+\left(1+\frac{2MG}{4\hat{r}}\right)^4(d\hat{r}^2+\hat{r}^2 d\Omega^2),
\end{equation}
with its corresponding metric components
\begin{equation}\label{eq:g00_conf}
    g_{00}=-\left(\frac{1-\frac{2MG}{4\hat{r}}}{1+\frac{2MG}{4\hat{r}}}\right)^2
\end{equation}
and
\begin{equation}\label{eq:gii_conf}
\begin{split}
    &g_{11}=\left(1+\frac{2MG}{4\hat{r}}\right)^4,\quad
    g_{22}=\left(1+\frac{2MG}{4\hat{r}}\right)^4r^2,\quad\\
    &g_{33}=\left(1+\frac{2MG}{4\hat{r}}\right)^4r^2\sin^2\theta.
\end{split}
\end{equation}
\end{subequations}
and the equivalent medium parameters obtained by inserting~\eqref{eq:g00_conf} and~\eqref{eq:gii_conf} into~\eqref{eq:Grav_Analog}
\begin{subequations}\label{eq:ref_BH}
\begin{equation}\label{eq:eps_mu_BH}
   \epsilon_{\mathrm{H}}=\mu_{\mathrm{H}}=\frac{(1+\frac{2MG}{4\hat{r}})^3}{1-\frac{2MG}{4\hat{r}}},
\end{equation}
\begin{equation}\label{eq:chi_BH}
    \rchi_{\mathrm{H}}=0.
\end{equation}
\end{subequations}
%

For 2D, we transform the coordinates from spherical ($\hat{r}, \theta, \phi$) to polar ($\rho, \phi$). Now given the circular symmetry of the medium parameter in the new coordinates, we choose a circularly-symmetric modulation. Therefore, the modulation has to move in radial direction ($\rho$).

For modulation with radial velocity ($\beta=\beta_\rho$) and radial acceleration ($a'=a'_\rho$) in the $z-x$ plane the averaged constitutive parameters can be derived in the same manner as~\eqref{eq:eps_tan_avg_mod} to~\eqref{eq:chi_avg_mod} by applying the continuity on the field components of $E_y$, $E_\phi$, $H_y$, $H_\phi$, $D_r$ and $B_r$, namely
\begin{subequations}\label{eq:mod_radial}
 \begin{equation}
     \ov\epsilon_y=\ov\epsilon_\phi=\ov\epsilon_{\|}=\frac{\Sigma_\epsilon-\beta^2\epsilon_1\epsilon_2\Sigma_\mu}{1-\beta^2\Sigma_\epsilon\Sigma_\mu},
 \end{equation}
  \begin{equation}
	\ov\epsilon_\rho=\ov\epsilon_{\perp}=\frac{2\epsilon_1\epsilon_2}{\Sigma_\epsilon},
 \end{equation}
  \begin{equation}
     \ov\mu_y=\ov\mu_\phi=\ov\mu_{\|}=\frac{\Sigma_\mu-\beta^2\mu_1\mu_2\Sigma_\epsilon}{1-\beta^2\Sigma_\epsilon\Sigma_\mu},
 \end{equation}
 \begin{equation}
	\ov\mu_\rho=\ov\mu_{\perp}=\frac{2\mu_1\mu_2}{\Sigma_\mu}
 \end{equation}
 and
   \begin{equation}
    \ov\rchi_\rho=\ov{\rchi}=\beta\frac{\Delta_\epsilon\Delta_\mu}{1-\beta^2\Sigma_\epsilon\Sigma_\mu},
 \end{equation}
 \end{subequations}
where $\Sigma_\epsilon=(\epsilon_1+\epsilon_2)/2$, $\Delta_{\epsilon}=(\epsilon_1-\epsilon_2)/2$ and similarly for $\Sigma_\mu$, $\Delta_\mu$.

Due to the time-reversal symmetry of the metric in~\eqref{eq:sti_conf} ($g_{0j}=0$) the equivalent bianisotropy term in~\eqref{eq:Grav_Analog_chi}($\rchi_j$) is zero. To have zero bianisotropy we need to have either the permittivity or the permeability modulated. Given the assumption of s polarization (electric field only along $y$) and the isotropic nature of the metric in the polar coordinates we only modulate the permittivity, and leave the permeability unmodulated ($\ov\mu=\mu_1=\mu_2$)
Now we find the $a'$, $\beta_0$ and $\epsilon_{1,2}$ that satisfy the following equation
 \begin{equation}\label{eq:eqv_BH}
	 \sqrt{\ov\epsilon_{\|}\ov\mu}=\sqrt{\epsilon_\mathrm{H}\mu_\mathrm{H}}.
 \end{equation}
 This procedure is carried out numerically and the FDTD simulation results of such equivalent modulation parameters are shown in Fig.~5.
 
\section{Light Trajectories in Schwarzschild Gravity Analogs}\label{Sec:Trajectories}
 
\begin{figure}[H] 
  \centering
  \includegraphics[width=0.5\textwidth]{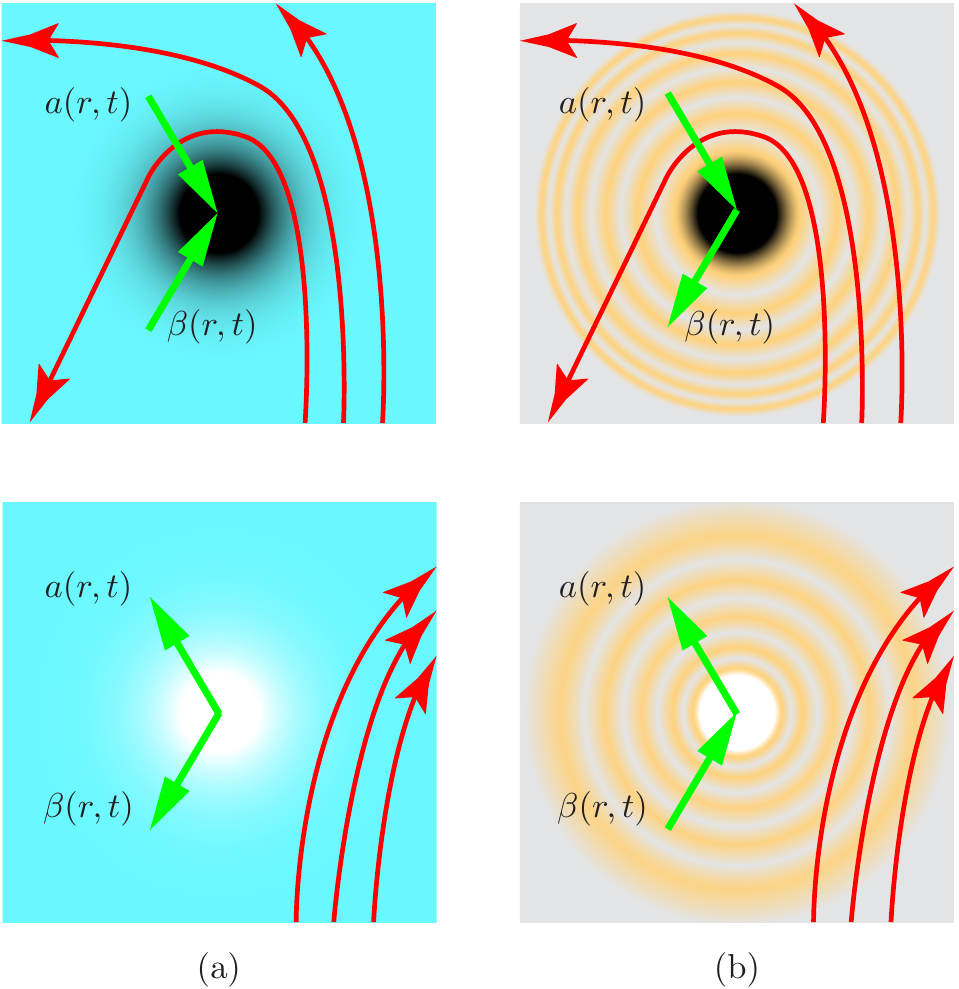}
  \caption{Qualitative light trajectories for Schwarzschild gravity analogs based on (a)~matter~\cite{schutzhold2002dielectric} and (b)~modulation, with centripetal acceleration, corresponding to a black holes (top panels), and centrifugal acceleration, corresponding to a white holes (bottom panels).}\label{fig:Qualtholes}
\end{figure}

Figure~\ref{fig:Qualtholes} depicts the qualitative trajectories of light near Schwarzschild gravity analogs, with Fig.~\ref{fig:Qualtholes}(a) corresponding to accelerated matter and Fig.~\ref{fig:Qualtholes}(b) corresponding to accelerated modulation.
\bibliography{Ref_Main_PRL}
\newpage
\clearpage
\end{document}